\documentclass[aps,pra,twocolumn,showpacs]{revtex4}
\usepackage{amsmath}
\begin{document}
\title{Scaling issues in ensemble implementations 
of the Deutsch-Jozsa algorithm}
\author{Arvind}
\thanks{Permanent Address: Department of Physics, 
Guru Nanak Dev
University, Amritsar  143005, INDIA}
\email{xarvind@andrew.cmu.edu}
\affiliation{Department of Physics, 
Carnegie Mellon University, Pittsburgh, PA
15213, USA}
\author{David Collins}
\thanks{Present Address: Department of Physics, 
Bucknell University, Lewisburg, PA 17837, USA}
\email{dcollins@bucknell.edu}
\affiliation{Department of Physics, 
Carnegie Mellon University, Pittsburgh, PA
15213, USA}
\begin{abstract}
We discuss the ensemble  version of the Deutsch-Jozsa (DJ)
algorithm which  attempts to provide a ``scalable''
implementation on an expectation-value NMR quantum computer.
We show that this ensemble implementation of the DJ
algorithm is at best as efficient as the classical random
algorithm. As soon as any attempt is made to classify all
possible functions with certainty, the implementation
requires an exponentially large number of molecules.  The
discrepancies arise out of the interpretation of mixed state
density matrices.
\end{abstract}
\pacs{ 03.67.Lx}
\maketitle
Conventional NMR implementations of quantum computing
algorithms require the preparation of pseudopure
states~\cite{knill97,chuang98a,cory98,schulman99,boykin02}.
There have been a few proposals to efficiently implement the
Deutsch Jozsa (DJ) algorithm on an NMR quantum information
processor using highly mixed
states~\cite{woodward00,myers02}.  The basic idea in these
ensemble schemes is to avoid pseudopure state preparation
which would require exponential resources and instead work
with highly mixed states close to thermal equilibrium.
These schemes need to be carefuly examined for their
`quantum character' and their efficiencies compared to
classical random algorithms.  In this paper we show that 
for the DJ problem, a parallel can be drawn between these
ensemble implementations and classical random algorithms.

We begin with a brief recapitulation of the DJ problem.
Consider the set of functions $f: \{0,1\}^n
\,{\longrightarrow}\, \{0, 1\}$.  If all $2^n\/$ inputs map
to the same output then the function is `constant' and if
half the outputs map to $0\/$ while the other half to $1\/$
then the function is `balanced'.  Functions which are
neither `constant' nor `balanced' are not considered here.
The task here is to determine the constant or balanced
nature of a given function.  Given an oracle which evaluates
$f(x)\/$ at an input $x \in \{0,1\}^n\/$, no deterministic
classical algorithm can carry out such a classification with
certainty without using at least $2^{n-1}+1\/$ invocations
of the oracle.  The quantum DJ algorithm on the other hand
accomplishes the classification task by invoking the oracle
only once ~\cite{cleve98,deutsch92}.
This it does by using a quantum oracle defined through the
unitary transformation on an $n\/$ qubit argument $\vert x
\rangle\/$ and a one qubit target $\vert y \rangle\/$,
\begin{equation}
 \vert x\rangle \vert y \rangle
\stackrel{\hat{U}_f}{\longrightarrow} \vert x\rangle\vert y
\oplus f(x) \rangle.  \label{q-oracle} 
\end{equation}
If a query to the oracle is assumed to come at a unit cost
then the quantum DJ algorithm provides an exponential
speedup over its classical counterpart.  The $n+1\/$ qubits
need to be in a pure quantum state for the algorithm to
work.  In liquid state NMR at room temperatures, which is
the most successful implementation of quantum information
processing to date, the quantum states of the spins are far
from pure. Therefore, to emulate the standard version of the
DJ algorithm, one has to prepare the system in a special
`pseudopure' state where the ensemble is divided into two
parts: a small subset in a given pure state, and the rest
acting as a uniform background with no contribution to the
signal.  However, the present preparation schemes for such
states lead to an exponential loss of signal because the
subset of spins which can be prepared in a `pure' state
decreases exponentially with the number of
qubits~\cite{warren97,boykin02}.

Recently, alternative schemes have been proposed to
circumvent this difficulty~\cite{woodward00,myers02}. These
effectively use  a computer with $n+1\/$ qubits, where the
first $n\/$ qubits are represented by a density matrix
$\frac{1}{2}I\/$ (a fully mixed state) and the last qubit is
in the pure state $\vert 0 \rangle$. The initial state thus
is
\begin{equation}
\rho_{\rm in}=\frac{1}{2^n} I \otimes I \otimes
\cdots I\otimes \vert 0 \rangle \langle 0 \vert 
\label{input_density1}
\end{equation}
and rewriting it in the computational basis 
$\{ \vert x \rangle \vert 0 \rangle \,|\, x=0 \cdots x=2^n-1\}$
gives
\begin{equation} 
\rho_{\rm in}=\frac{1}{2^n}\sum_{x=0}^{2^n-1}
\vert x \rangle \langle x \vert \otimes \vert 0 \rangle
\langle 0 \vert 
\label{input_density2} 
\end{equation}
This preparation is followed by the standard quantum
oracle query described in Eq.~(\ref{q-oracle}), yielding
\begin{equation} 
\rho_{\rm
out}=\frac{1}{2^n}\sum_{x=0}^{2^n-1} \vert x \rangle \langle x
\vert \otimes \vert f(x) \rangle \langle f(x) \vert
\label{output} 
\end{equation}
Before we extract the constant or balanced nature of the
function we note that this is not an entangled state.  The
entanglement is missing because of the special choice of the
initial state.  As a matter of fact the oracle is capable of
generating entanglement and the standard pure state version
of the quantum algorithm relies on
entanglement~\cite{cleve98,deutsch92}.

The information about the function is contained entirely in
the target qubit whose reduced density matrix is
\begin{equation}
\rho_{\rm target} =
\frac{1}{2^n}\sum_{x=0}^{2^n-1}\vert f(x) \rangle \langle f(x)
\vert . 
\label{target-density}
\end{equation}
The expectation value  of $\sigma_z\/$ in this state will
immediately reveal the `constant' or `balanced' nature of
the function $f\/$, with the result: 
\begin{equation}
\langle  \sigma_z \rangle = 
\left\{\begin{array}{ll}\pm 1\quad &{\rm Constant} \quad f\\ 
0 \quad &{\rm Balanced} \quad f
\end{array} \right.
\label{measurement} 
\end{equation}
By actually carrying out such a measurement the function can
be classified with a single invocation of the quantum oracle
without the associated problems of preparing pure or
pseudo-pure states. This scheme is particularly suitable for
implementation on an NMR quantum information processor where
the thermal equilibrium state can be easily transformed into
the maximally mixed state of Eq.~(\ref{input_density1}) and
expectation value measurements are natural.

For comparison we describe a classical scenario which in
essence mimics the `quantum' scheme described above.
Instead of the NMR qubits, consider classical bit strings of
length $n+1\/$.  Further, assume that we have $2^n\/$ such
strings and each string is in a different state for the
first $n\/$ bits, thereby providing representation to all
possible states of the first $n$-bits. The $(n+1)$th bit in
each string is set to `0' and acts as the target bit.  Now
the application of the classical oracle ($x \longrightarrow
f(x)\/$ with $f(x)\/$ appearing on the target)  to all the
copies will yield the function values at all the $2^n\/$
input points and this value will be stored in the target bit
in each copy. The constant or balanced nature of the
function can then be obtained by  adding these values by
appropriate gates.  The values add to $2^n\/$ or $0\/$ for a
constant function and to $2^n/2\/$ for a balanced function.
This is analogous to the case of the expectation value
quantum algorithm using maximally mixed states. In our view
this scheme is thus fully classical, using separable states
at all stages and camouflaged in the language of quantum
mechanics.  The exponential resource is explicit in the
classical situation while it is hidden in the definition of
the input density matrix of Eq.~(\ref{input_density1}) for
the ensemble quantum case.  The deceptively simple fact that
one has effectively prepared the state of the first $n\/$
qubits in the density matrix $\frac{\displaystyle
1}{\displaystyle 2^n} \left(I\otimes I \otimes I ... \otimes
I\right)\/$ requires that we have at least $2^n$ molecules!

Every density operator can be viewed as an ensemble of pure
states occurring with certain probabilities. The existence of
a decomposition
\begin{equation} 
\rho = \sum_i p_i \vert \psi_i\rangle \langle
\psi_i \vert \quad {\rm with} \quad p_i\geq 0, \quad \sum p_i =1
\label{ensemble}
\end{equation}
implies an ensemble $ \left\{p_i, \,\vert \psi_i\rangle
\right\}$ for $\rho$, where the state $\vert \psi_i
\rangle\/$ occurs with probability $p_i$. It is to be noted
that the states $\vert \psi_i\rangle\/$  need not be
orthogonal and the decomposition given above is not unique
for mixed states.  However, whatever can be determined from
$\rho\/$ can be consistently thought of as deriving from any
one of the ensembles.  The ensemble scheme that culminates
in Eq.~(\ref{measurement}) is illustrated by considering the
situation in which each ensemble member is taken to be in
one of the computational basis (pure) states, $\left\{ \vert
x \rangle \vert 0 \rangle \; | \; x=0 \ldots 2^n-1
\right\}$, each of which occurs with probability $p_x =
1/2^n.$ This corresponds to the input density operator of
Eq.~(\ref{input_density2}).  Now view the scheme as it is
applied on each member of the ensemble, which means that
each member of the ensemble independently computes the
function on the input state of that member and the output
appears on the $(n+1)$th qubit of each member molecule.  The
measurement of the average of $\sigma_z \/$ then ostensibly
reveals the constant or balanced nature of the function as
described in Eq.~(\ref{measurement}).

Now imagine that one chooses to work with a fixed number of
molecules $M\/$. As the number of qubits $n\/$ increases,
soon one will reach a stage when  $M < 2^n/2\/$. In this
case more than half  the computational basis states cannot
find representation in the ensemble because there are simply
not enough molecules! Therefore there are  always  balanced
functions which  will have the same value over these $M\/$
states and will get classified as constant, despite assuming
the best situation, where all the molecules are assumed to
be in different states.  We will see later that the actual
scheme is even more inefficient because all the molecules
cannot be in different computational basis states.  Thus,
for the scheme to work for all functions one needs at least
$2^n/2\/$ molecules in the ensemble, a number which grows
exponentially with $n$. In other words, when the number of
molecules is smaller than $2^n\/$, there is no way one can
prepare the input density matrix of
Eq.~(\ref{input_density1}).  We will return to this point
later.

How many balanced functions escape classification for a
given $M$?  Assume $N=2^n$ is the input set size. The number
of constant functions is $2$ which is independent of $n\/$
while the number of balanced function is $\,{}^NC_{N/2}$.
If the  scheme is used with a number of molecules $1\leq
M\leq \frac{N}{2}\/$, the balanced functions which escape
classification are the ones which have same value ($0\/$ or
$1\/$) for the first $M\/$ inputs.  The number of functions
which have the value $1\/$ ($0$) for the first $M\/$ inputs
is the same as the number of ways one can distribute the
remaining $ N/2-M\,\,$ $1$'s ($0$'s) on $N-M$ inputs,
giving:
\begin{equation}
\begin{array}{r}{\rm No.~of~Balanced~Fns.}\\ {\rm Classified~as
~Constant} \end{array} \,=\,
2 {N-M \choose N/2-M}
\end{equation}
Dividing this by the total number of balanced functions
gives the fraction of balanced functions for which the
schemes fails
\begin{equation}
{\rm Failure~Fraction}=2 {N-M \choose N/2-M}/ {N \choose N/2}
\label{fail-fraction}
\end{equation}
This fraction diminishes quite fast as $M\/$ increases from
$0\/$ toward $\frac{N}{2}\/$, increasing the efficiency of
the algorithm.  The fact that for most cases the scheme will
work with a relatively small number of molecules has nothing
to do with quantum mechanics. The classical randomized
algorithm too will work to the same extent. In fact the
above counting is valid for the classical algorithm as well.
It is well known that there is an efficient randomized
classical algorithm  for the DJ problem and we conclude that
the expectation value ensemble scheme in the best case is
equivalent to it~\cite{brazier03}.

Even if  we use a molecule number $M>2^n\/$ how safely can
we say that the initial maximally mixed state has been
realized and we are able to classify all functions with
certainty?  It is possible that, in a given experimental
implementation of the algorithm, all ensemble members are in
the same initial state $\vert x^\prime \rangle \vert 0
\rangle.$ In this event, the algorithm only evaluates
$f(x^\prime)$ and the measurement outcome will be that for a
constant function regardless of the nature of $f$. Thus, in
contrast to the conventional Deutsch-Jozsa algorithm, there
can be no way of determining the function type with
certainty. In this sense, this ensemble algorithm for
solving the Deutsch-Jozsa problem is not deterministic and
must be compared to probabilistic classical algorithms. We
shall consider the probability with which each correctly
determines the function type and show that, regardless of
the ensemble size, the standard probabilistic classical
algorithm is superior to this ensemble quantum algorithm.

Suppose that the ensemble consists of $M$
identical, independent $n+1$ qubit molecules. Each member
of the ensemble will be subject to the unitary of
Eq.~(\ref{q-oracle}), which can be re-expressed as
\begin{equation} 
\hat{U}_f = \hat{P}_0(f) \otimes I +
\hat{P}_1(f) \otimes \sigma_x \label{new-oracle}
\end{equation}
where 
\begin{eqnarray}
\hat{P}_0(f) & := & \sum_{x: f(x)=0} 
\vert x\rangle \langle x \vert
\nonumber \\
\hat{P}_1(f) & := & \sum_{x: f(x)=1} 
\vert x\rangle \langle x \vert   
\end{eqnarray}
project onto subspaces of $n\/$ qubit argument while $I\/$
and $\sigma_x\/$ act on the target qubit. 
For the target qubit, $\langle \sigma_z \rangle$ is
approximated by 
\begin{equation}
\overline{z} := \frac{1}{M}
\sum_{j=1}^M z_j
\end{equation}
where $z_j=\pm 1$ are the outcomes of projective measurement
($z_j=1$ corresponding to $\vert 0\rangle \langle 0\vert$
and $z_j = -1$ to $\vert 1\rangle \langle 1\vert$) on the
target qubit for individual ensemble members.

The only assumption that we make about the ensemble members'
initial states is that they occur with probabilities
described by the density operator of
Eq.~(\ref{input_density2}). Then, for any ensemble member
$j$,
\begin{subequations}
\begin{eqnarray}
{\rm Prob}(z_j=+1 \; | \; f) & = & \textrm{Tr}_{\rm arg}
                                   \left( \hat{P}_0(f) 
                                   \rho_{\rm in}  
                                   \right) \\
\textrm{Prob}(z_j=-1 \; | \; f) & = & \textrm{Tr}_{\rm arg}
                                   \left( \hat{P}_1(f) 
                                   \rho_{\rm in}  
                                   \right)
\end{eqnarray}
\end{subequations}
where measurements are performed immediately after algorithm
unitaries and the traces are taken over the argument
register only. Note that each constant function yields one
measurement outcome with certainty: $z_j = +1$ for $f=0$ and
$z_j = -1$ for $f=1.$ Thus $\overline{z} = +1$ for $f=0$ and
$\overline{z} = -1$ for $f=0$. Whenever $\overline{z}$
departs from $\pm1$ it is clear that $f$ is balanced.
However, the extent to which such a departure is noticeable
depends on the available measurement resolution, which can
be expressed in terms of outcomes of function register
measurements on individual ensemble members.  Suppose that
it is possible to distinguish two ensemble averages only
when they differ in $R$ (out of $M$) or more individual
measurement outcomes. Then we regard two ensemble averages
as distinct provided that $\left|
\overline{z}-\overline{z}^\prime \right| \geq R/M.$ This
motivates the following protocol for deciding the algorithm
outcome:
\begin{subequations}
\begin{eqnarray}
\overline{z} \geq 1-R/M & \Rightarrow & f = 0 \\
\overline{z} \leq -1 +R/M & \Rightarrow & f = 1 \\
1-R/M > \overline{z} > -1 + R/M & \Rightarrow & f \;
\textrm{balanced}.
\end{eqnarray}
\end{subequations}
The issue is to determine the probability with which this
protocol will correctly identify the function type. Constant
functions will always be identified correctly and we need
only to find the probability that a balanced function will
give $\overline{z} \geq 1-R/M$ or $\overline{z} \leq -1
+R/M$. These are the probabilities that a balanced function
will return $z_j = -1$ or $z_j = +1$ at most $R-1$ times
respectively. For a balanced function $\textrm{Prob}(z_j=+1)
= \textrm{Prob}(z_j=-1) = 1/2.$ The probability that we
incorrectly declare a balanced function to be constant is
\begin{eqnarray}
p_{\rm fail} & = & 2  \sum_{k=0}^{R-1} {M \choose k} 
\left( \frac{1}{2} \right)^k \left( \frac{1}{2} \right)^{M-k} 
\nonumber \\
& = & \frac{1}{2^{M-1}} \sum_{k=0}^{R-1} {M \choose k}.
\label{ensemble-fail-prob}
\end{eqnarray}
In the best conceivable case $R=1,$ giving 
$p_{\rm fail} = 1/2^{M-1}.$

To account for the spatial resources offered by the ensemble
we consider the application of $\hat{U}_f$ to the ensemble
containing $M$ members as equivalent to $M$ oracle calls. We
must then compare this ensemble algorithm to a classical
random algorithm that uses $M$ oracle calls. In the
classical random algorithm one begins by choosing $x_1$
randomly and evaluating $f(x_1)$. The next step is to choose
$x_2 \neq x_1$, evaluate $f(x_2)$ and compare the result to
$f(x_1).$ If the two differ then $f$ is balanced. If not
pick $x_3$ which differs from both $x_1$ and $x_2,$ and
compare $f(x_3)$ to $f(x_2)$ and $f(x_1)$, etc.... The
algorithm terminates when $f$ returns different outcomes or
has been evaluated on $2^n/2 +1$ different inputs.  The
classical random algorithm never misidentifies a constant
function and identifies a balanced function $f\/$ as
constant only when $f(x_1)=f(x_2)=\cdots =f(x_M)\/$. The
probability of failure is the probability with which this
occurs. The outcome $f(x_1)=0\/$ occurs with probability
$(N/2)/N\/$. Given $f(x_1)=0$, $f(x_2)=0$ occurs with
probability $(N/2-1)/(N-1)$.  Continuing, the probability
that $x_1, \ldots , x_M$ are all such that $f(x_k)=0$ is:
\begin{eqnarray}
p_{\rm classical}^0 &= &\frac{N/2}{N} \frac{N/2 -1}{N -1} 
\ldots \frac{N/2 -M+1}{N-M+1} \nonumber \\
& =& {N/2 \choose M}/ {N \choose M}
\end{eqnarray}
Similarly the probability $p_{\rm classical}^1$ that $x_1, 
\ldots , x_M$ are all
such that $f(x_k)=1$ can be computed and it turns out to be
same as $p_{\rm classical}^0$.
Thus the probability of failure is:
\begin{equation}
p^{\rm classical}_{\rm fail} = 2 {N/2 \choose M}/ {N \choose M}.
\end{equation}
As expected, this turns out to be same as the failure
fraction given in Eq.~(\ref{fail-fraction}) for the quantum
ensemble version with the unrealistic assumption that every
molecule is in a different computational basis state.
However, note that for $N/2 >k>0$, $(N/2-k)/(N-k) < 1/2,$
which implies that 
\begin{equation}
p^{\rm classical}_{\rm fail} < 2
\left( \frac{1}{2} \right)^M \leq  p_{\rm fail}.
\label{random-classical}
\end{equation}
Thus the probability of failure for the classical random
algorithm is strictly less than that of the ensemble quantum
version discussed here.

The central issue is therefore one of interpreting density
matrices. What is relevant here are the inferences that can
be drawn from measurement outcomes on  quantum systems whose
states are described by density matrices. In all cases the
density matrix merely provides the probability distribution
for outcomes of various measurements. The accuracy with
which such a distribution is realized improves with an
increasingly large ensemble.  Imagine a single quantum
system which is handed over to us with no information about
it. What quantum state or density matrix will we be able to
assign to it? To express our complete lack of information
about this system we have to assign equal weightage to all
possible outcomes in all bases and therefore a density
matrix proportional to identity is the  best  choice.  In
this extreme case, measurement yields one of all possible
outcomes and one cannot reliably infer anything from this.
The density operator merely reflects our lack of knowledge
in the state of the system. It is only when measurements are
performed on many copies described by the same density
operator that outcomes or more precisely, the average
outcome, carry any meaningful information. This is no more
than standard statistical sampling and for an $n\/$ qubit
density matrix proportional to the identity, the variance
scales as $2^{2n}\/$, indicating that one typically needs
$O(2^n)\/$ samples (measurement outcomes) to make sensible
inferences from measurements. Accordingly the ensemble size
would have to scale as $O(2^n)\/$ before we can consider
this density operator to have been realized accurately, at
least  in terms of measurement outcome averages. Here the
ensemble begins to appear as a collection of quantum systems
with states described according to the density matrix.  It
should be noted that, for this version of the DJ algorithm,
the situation is less dire since inferences are made from
measurements on the target qubit alone. Hence $n\/$ does not
appear in the failure probability in
Eqn.~(\ref{ensemble-fail-prob}).  However, as clear from
Eqn.~(\ref{random-classical}) a classical random algorithm
does the task better.

These ensemble computing ideas might work for other
algorithms and give a genuine speed up over classical or
classical random algorithms.  One possibility is efficient
simulation of quantum systems~\cite{knill98} and our result
does not pertain to this.  It is worthwhile to explore the
exact implications of this model which will be taken up
elsewhere.

\begin{acknowledgments} 
Numerous stimulating discussions with R.~B.~Griffiths and
Kavita~Dorai are acknowledged.  The research
effort is funded by the NSF Grant No.~0139974.
\end{acknowledgments}

\end{document}